\documentstyle[12pt]{article}

\newcommand{\beq}[1]{\begin{equation} \label{#1} }
\newcommand{\eeq}{\end{equation}}
\newcommand{\av}[1]{\langle #1 \rangle}

\begin{document}

\title{Prospects for time reversal invariance studies with the use of five-fold correlation}

\author{A. L. Barabanov$^1$, A. G. Beda$^2$, A. F. Volkov$^3$\\
{\normalsize\it $^1$ Kurchatov Institute, Moscow}\\
{\normalsize\it $^2$ Institute of theoretical and experimental physics, Moscow}\\
{\normalsize\it $^3$ Moscow institute of radiotechnique, electronics and automatics, Moscow}}

\date{}

\maketitle

\begin{abstract}

Advantages of investigation of time reversal invariance violation with the use of three-fold (P,T-odd) and five-fold (P-even,T-odd) correlations in the interaction of resonance neutrons with nuclei are briefly considered. Possible enhancements of T-odd effects in both cases are discussed. It is shown that the study of five-fold correlation is a perspective way to test time reversal invariance. Prospects to realize dynamical nuclear alignment method for measurements of five-fold correlation are described.

\end{abstract}
 
\section{Introduction}
 
Interaction of slow polarized neutrons with medium and heavy nuclei is a sensitive tool to study the fundamental symmetries (space parity and time reversal invariance) due to large en\-han\-ce\-ment of the effects of parity (P) violation and time reversal invariance  (T invariance) violation in compound resonances. The vast majority of  performed experimental works has been dedicated to investigation of P violation resulting from weak nucleon-nucleon interactions (PV interactions). Non-zero P odd effects have been observed for more than three tens of nuclei [1]. In these experiments a phenomenon of enhancement of the parity violation effects in compound resonances was clearly demonstrated. The measured effects range up to 10$^{-2}$ -- 10$^{-1}$ exceeding by several orders of magnitude the parity violation effects in nucleon-nucleon interaction that are of the order of
10$^{-8}$ -- 10$^{-7}$.

The  methods to test T invariance in the interaction of slow polarized neutrons with oriented nuclei were proposed in a number of works. Besides, an important hypothesis has been put forward in [2,3] that a mechanism of enhancement of T violation in compound resonances is the same as for P violation. Thus, due to large enhancement factors, methods to search for interactions violating T invariance (TV interactions) based on the use of slow neutrons are very promising. Real progress in this field is hampered by methodical problems as well as to some extent  by insufficient intensity of currently available sources of resonance neutrons.

In this work a current status of hypothetic TV interactions is discussed, and methods of their investigation in neutron-nuclei interaction are compared. We would like to show that at the moment a search for  so-called five-fold correlation in the total cross section of the interaction of resonance polarized neutrons with aligned nuclei, which is a promising way to obtain new information on TV interactions, may be practically realized. Note that the construction of two neutron spallation sources (JSS, Japan and SNS, USA) is now in progress. Their intensity will be five-ten times as great as the intensity of the existing neutron spallation sources.
 
\section{Three-fold and five-fold correlations in total cross section of the interaction of slow neutrons with nuclei}
 
Total cross section of the interaction of s-wave neutrons with target nuclei depends  on relative orientation of the spins ${\bf s}$ and ${\bf I}$ of neutrons and nuclei, respectively. A contribution of p-wave to the interaction amplitude is suppressed at low energy by the factor $kR\sim 10^{-3}$, where $k$ is the neutron wave number, and $R$ is the nucleus radius. However, a neutron-nucleus p-wave interaction is pronounced in p-wave resonances. Although the integral cross section within the limits of a p-wave resonance is suppressed by the factor $kR$ as compared to the typical integral cross section of the s-wave resonance, strong p-wave resonances are clearly prominent on the potential s-wave background.

The total cross section $\sigma_{tot}$ of neutron-nucleus interaction depends on relative orientation of the vectors ${\bf s}$, ${\bf I}$ and ${\bf k}$, where ${\bf k}$ is the neutron wave vector, if s- and p-waves are taken into account. Usually one measures the number $N({\bf s},{\bf I},{\bf k})$ of neutrons passing through oriented nuclear target, and a typical observable is the asymmetry
\beq{2.1}
\alpha=\frac{N_+-N_-}{N_++N_-},
\eeq
where $N_+$ and $N_-$ are numbers of detector counts for two opposite directions of the vector
${\bf s}$. In general, the total number of terms in $\sigma_{tot}$ depends on the type of orientation of target nuclei.

Let the axis $z$ is directed along the vector ${\bf I}$, i.e. along the orientation axis of an ensemble of target nuclei, and $m$ is the projection of spin $I$ of a nucleus on the axis $z$. An ensemble of nuclei is oriented if populations of substates with projections $m$ are different.

An ensemble of nuclei is said to be polarized if the parameter of polarization
\beq{2.2}
p_1(I)=\frac{\av{m}}{I}
\eeq
is not zero, where $\av{m}$ is the average projection for the ensemble of nuclei.

An ensemble of nuclei is said to be pure aligned if $p_1(I)=0$ but the parameter of alignment
\beq{2.3}
p_2(I)=\frac{3\av{m^2}-I(I+1)}{I(2I-1)}
\eeq
is not zero. Only nuclei with spin $I>1/2$ can be aligned. At pure alignment, the populations of substates with projections $m$ and $-m$ are equal (as the result $\av{m}=0$ and $p_1(I)=0$).

In general case, both introduced parameters may be non-zero. Then one speaks about oriented nuclei characterized both by the polarization $p_1(I)$ and the alignment $p_2(I)$.

With account for polarization and alignment of target nuclei, the total cross section $\sigma_{tot}$ includes 8 terms differently depending on relative orientation of the vectors ${\bf s}$, ${\bf I}$ and ${\bf k}$. All these terms are presented and discussed, e.g., in [4]. Investigations of P violation in the interaction of slow neutrons with nuclei are based mainly on the correlation $({\bf s}{\bf k})$, which is proportional to neutron polarization but needs no nuclear orientation. Investigations of TV interactions are possible due to T non-invariant three-fold $({\bf s}[{\bf k}\times{\bf I}])$ and
five-fold $({\bf s}[{\bf k}\times{\bf I}])({\bf k}{\bf I})$ correlations. Both these correlations are proportional to neutron polarization, but three-fold correlation is additionally proportional to nuclear polarization while five-fold correlation is proportional to nuclear alignment.

In the work [5], which was among the first dedicated to five-fold correlation in neutron-nucleus interaction, it was pointed out in particular that these T non-invariant correlations differ by space parity (the correlation $({\bf s}[{\bf k}\times{\bf I}])$ is P odd, while the correlation
$({\bf s}[{\bf k}\times{\bf I}])({\bf k}{\bf I})$ is P even). Therefore, they are sensitive to different type of fundamental TV interactions. One distinguishes PVTV interactions violating both T and P invariances and PCTV interactions violating T invariance but conserving P invariance (see, e.g., [6]).

Thus, investigations of three-fold and five-fold correlations are not competitive ways to search for fundamental TV forces, but are complement experiments which may give information on TV interactions of different nature, PVTV and PCTV, respectively.

\section{Current status of PVTV and PCTV interactions}
 
Present models of the interactions violating T invariance were considered in [6]. In the framework of Standard Model (SM) PVTV forces that may substantially manifest themselves in neutron-nucleus interaction result from $\theta$-term in QCD Lagrangian. There are many assumptions about hypothetic PVTV forces beyond SM.

Effective PCTV interactions may be obtained in the framework of SM from $\theta$-term and usual weak PV forces, but in this case the observables are so small that they cannot be measured. Thus, in nuclear experiments one searches for PCTV interactions beyond SM. There are several ways to introduce such PCTV forces.

Measurements of electric dipole moments (EDM) of neutron and some atoms give the best limitation for PVTV interactions. Let
\beq{3.1} 
\lambda_{PVTV}=\frac{\av{V_{PVTV}}_{s.p.}}{\av{V_{PV}}_{s.p.}}
\eeq
is a ratio of typical single particle matrix elements of PVTV and PV interactions, $V_{PVTV}$ and $V_{PV}$, respectively. The limit
\beq{3.2}
\lambda_{PVTV} < 10^{-4}
\eeq
was obtained in [6] from the EDM data.

The situation with limit on PCTV interactions is more intricate. The PCTV forces alone cannot induce EDM because they are P even, while EDM is not only T non-invariant but P odd as well. However, PCTV forces together with weak PV interactions may involve EDM. Therefore, limit on PCTV forces may be obtained from the EDM data but only in the framework of some model assumptions about PV interactions at small distances.

In 1990s there was a confidence that the data on EDM of neutron and atoms put more strict limits for PCTV interactions than direct searches for T non-invariant, P even effects, e.g., test of detail balance in nuclear reactions. Let
\beq{3.3}
\lambda_{PCTV}=\frac{\av{V_{PCTV}}_{s.p.}}{\av{V}_{s.p.}}
\eeq
is the ratio of typical single particle matrix elements of PCTV interaction $V_{PCTV}$ and strong interaction $V$. Note, that it is the practice to compare PCTV interactions just with strong interaction $V$ because both are P even. Summing the results of theoretical works devoted to relations between PCTV forces and EDM, a rather strict limit
\beq{3.4}
\lambda_{PCTV} < 10^{-8}
\eeq
was obtained in [6].

Last years this limit was reconsidered in [7]. More exactly, it was pointed out in [7] that the strict limitation on $\lambda_{PCTV}$ from EDM data arises in the framework of a certain assumption ("Scenario A"). Really, there are no fundamental reasons to follow this assumption. In the framework of an alternative approach ("Scenario B") the EDM data give no definite value for $\lambda_{PCTV}$. In this case one may use only a relatively weak limit
\beq{3.5}
\lambda_{PCTV} < 10^{-4}
\eeq
resulting from direct searches for T non-invariant, P even effects. Among them is the experiment on measurement of five-fold correlation $({\bf s}[{\bf k}\times{\bf I}])({\bf k}{\bf I})$ in the interaction of fast neutrons with aligned nuclei $^{165}$Ho [8].
 
\section{Enhancement of PVTV and PCTV effects in a p-wave compound resonance}

Let $\alpha_{PV}$, $\alpha_{PVTV}$ and $\alpha_{PCTV}$ are measurable asymmetries in the experiments on searches for P violation (correlation $({\bf s}{\bf k})$\,), simultaneous violation of P and T invariance (three-fold correlation $({\bf s}[{\bf k}\times{\bf I}])$\,) and T violation with P conservation (five-fold correlation\linebreak
$({\bf s}[{\bf k}\times{\bf I}])({\bf k}{\bf I})$\,), respectively. The magnitude of P odd effect in a p-wave resonance is determined by (see, e.g., [9])
\beq{4.1}
\alpha_{PV}\sim\frac{g^n_s}{g^n_p}\,\frac{\av{V_{PV}}_c}{D_c},
\eeq
where $g^n_p$ and $g^n_s$ are the neutron amplitudes of p-wave and s-wave resonances mixed by the weak PV interaction, $\av{V_{PV}}_c$ is the matrix element of PV interaction between s- and p- resonances states, and $D_c$ is the distance between the resonances. Let us emphasize that beyond p-wave resonance the measured asymmetry $\alpha_{PV}$ is much smaller than the given estimation. Large enhancement of the P odd effect in a p-wave resonance is known as resonance enhancement.

The P odd asymmetry in a p-wave resonance is proportional to the ratio
$g^n_s/g^n_p\sim (kR)^{-1}\sim 10^3$ which is usually called the factor of structural enhancement. The ratio $\av{V_{PV}}_c/D_c$ is related to the analogous ratio of single particle quantities as follows
\beq{4.2}
\frac{\av{V_{PV}}_c}{D_c}\sim
\sqrt{N_c}\,\,\frac{\av{V_{PV}}_{s.p.}}{D_{s.p.}},
\eeq
where $N_c\sim 10^6$ is the number of single particle components in the wave function of compound nucleus. The factor $\sqrt{N_c}\sim 10^3$ is called the factor of dynamical enhancement. Here $\av{V_{PV}}_{s.p.}/D_{s.p.}\sim Gm^2_{\pi}\sim 10^{-7}$ is an estimation of P odd effect in nucleon-nucleon interaction.

Thus, the total enhancement (structural and dynamical) of P odd effect in a p-wave resonance as compared to P odd effect in nucleon-nucleon interaction may be as great as 10$^6$ and the effect may reach the value of 10$^{-1}$
\beq{4.3}
\alpha_{PV}\sim (kR)^{-1}\sqrt{N_c}\,\,Gm^2_{\pi}\sim 10^{-1}.
\eeq
It is of interest that just such large P odd effect has been observed in the p-wave resonance of the nucleus $^{139}$La, namely $\alpha_{PV}\sim 10^{-1}$ [1]. Effects of the order of
10$^{-3}$ -- 10$^{-2}$, i.e. the enhancement factors 10$^4$ -- 10$^5$, are more typical.

Let us now consider PVTV asymmetry related to three-fold correlation
$({\bf s}[{\bf k}\times{\bf I}])$. In a p-wave resonance one gets [9]
\beq{4.4}
\alpha_{PVTV}\sim\frac{g^n_s}{g^n_p}\,\frac{\av{V_{PVTV}}_c}{D_c}.
\eeq
This estimate involves both structural and dynamical enhancement factors.

Using the quantity $\lambda_{PVTV}$ (\ref{3.1}), one obtains
\beq{4.5}
\alpha_{PVTV}\sim\lambda_{PVTV}\,\,\alpha_{PV}.
\eeq
Taking into account that $\lambda_{PVTV} < 10^{-4}$ and $\alpha_{PV}\sim 10^{-1}$ in the p-wave resonance of the nucleus $^{139}$La, we conclude that an experiment on the search for PVTV forces in the transmission of slow polarized neutrons through polarized target $^{139}$La in the same p-wave resonance should be performed with the accuracy
\beq{4.6}
\alpha_{PVTV} < 10^{-5}.
\eeq
Only at such accuracy the study of three-fold correlation $({\bf s}[{\bf k}\times{\bf I}])$ allows to exceed the limitation on PVTV forces resulting from the present EDM data.

Now let us discuss PCTV asymmetry, related to five-fold correlation
$({\bf s}[{\bf k}\times{\bf I}])({\bf k}{\bf I})$. It was shown in [3] that in a p-wave resonance the effect is
\beq{4.7}
\alpha_{PCTV}\sim\frac{\av{V_{P'TV}}_c}{D_c}.
\eeq
Since PCTV interaction mixes resonances with the same parity (in the case under consideration -- p-wave resonances), here the factor of structural enhancement is absent. However, due to the relationship
\beq{4.8}
\frac{\av{V_{PCTV}}_c}{D_c}\sim
\sqrt{N_c}\,\,\frac{\av{V_{PCTV}}_{s.p.}}{D_{s.p.}}
\eeq
the dynamical enhancement expressed by the factor $\sqrt{N_c}\sim 10^3$ takes place.

Using the quantity $\lambda_{PCTV}$ (\ref{3.3}) and taking into account that a typical single particle matrix element of strong interaction $\av{V}_{s.p.}$ is of the scale of single particle distance between the levels $D_{s.p.}$, one gets for PCTV asymmetry in a p-wave resonance
\beq{4.9}
\alpha_{PCTV}\sim\sqrt{N_c}\,\,\lambda_{PCTV}.
\eeq
Thus, due to dynamical enhancement one may expect that PCTV asymmetry in a p-wave resonance of heavy nucleus is increased to $\sqrt{N_c}\sim 10^3$ times as compared to possible PCTV effect in a few nucleon system which magnitude is given by the factor $\lambda_{PCTV}$. 

It was pointed out in [7] that hadron reactions give the limit: $\lambda_{PCTV} < 10^{-4}$. Hence it follows that the study of five-fold correlation $({\bf s}[{\bf k}\times{\bf I}])({\bf k}{\bf I})$ allows to exceed this limitation on PCTV forces provided the experiment will be performed with the accuracy
\beq{4.10}
\alpha_{PCTV} < 10^{-1}.
\eeq

\section{Methodical problems of measurements of three-fold and five-fold correlations}

It is well known that the measurement of three-fold correlation $({\bf s}[{\bf k}\times{\bf I}])$ in the interaction of slow neutrons with nuclei is hampered by severe methodical problems. One should measure the asymmetry $\alpha_{PVTV}$ of neutron transmission through polarized nuclear target at two opposite directions of neutron spin ${\bf s}$, namely along and opposite the vector
$[{\bf k}\times{\bf I}]$, where the vector ${\bf I}$ is normal to the vector ${\bf k}$.

To polarize nuclear target one usually uses a strong magnetic field ${\bf H}$, therewith the vector ${\bf I}$ is directed along ${\bf H}$. It causes a Larmor precession of neutron spin ${\bf s}$ in the target. This precession results in non-zero neutron helicity and the P odd effect due to correlation $({\bf s}{\bf k})$. Since P-odd effect is at least 3-4 order greater than T non-invariant, P odd effect, the orthogonality of the vectors ${\bf s}$ and ${\bf k}$ during the measurement should be provided with the accuracy better than 10$^{-5}$, that is a very difficult problem. Besides, even without magnetic field the problem of precession of the vector ${\bf s}$ around the vector ${\bf I}$ remains because it additionally results from nuclear pseudomagnetism, i.e. from the correlation
$({\bf s}{\bf I})$. This problem as well as possible methods to eliminate false effects  are now under discussion.

In the case of five-fold correlation $({\bf s}[{\bf k}\times{\bf I}])({\bf k}{\bf I})$ one should measure the asymmetry $\alpha_{PCTV}$ of neutron transmission through an aligned nuclear target at two opposite directions of neutron spin ${\bf s}$, namely along and opposite the vector
$[{\bf k}\times{\bf I}]$, therewith the vector ${\bf I}$ should be directed at some angle (e.g. $\pi/4$) to the vector ${\bf k}$.

It is of importance that there is no need in magnetic field to align the nuclei. Besides, nuclear polarization is equal to zero at pure alignment. Thus, there is no pseudomagnetic interaction of neutrons with aligned nuclear target and, therefore, there is no neutron spin precession in the target due to pseudomagnetism.

One can note that even in the presence of some  precession the corresponding problems in the case of five-fold correlation are not nearly so serious as in the case of three-fold one. Indeed, the above estimated measurement accuracy for PCTV asymmetry $\alpha_{PCTV} < 10^{-1}$ is of the same order as a possible P odd effect $\alpha_{PV}$. Therefore, the orthogonality of the vectors
${\bf s}$ and ${\bf k}$ during the experiment should be provided with the accuracy 10$^{-2}$ that is not the problem.

Up to now the progress in studies of T violation with the use of five-fold correlation was hampered mainly by a deficit of aligned targets. Recently in [4] a new method of dynamical nuclear alignment (DNA) was proposed. It can allow to increase substantially the number of aligned nuclei available for experiments.

\section{Prospects of realization of DNA method}

DNA method proposed in [4] is analogous to the well-known method of dynamical nuclear polarization (DNP) [10] but DNA does not require an external magnetic field. One widely uses DNP to produce polarized proton targets. 

The ground state of a nucleus with spin I is split in the magnetic field into $(2I+1)$ magnetic substates (at DNP). At DNA the ground state of a nucleus with spin $I>1/2$ is split in a crystal due to an interaction of quadrupole nuclear moment with a gradient of the crystal electric field (for instance, at I= 3/2 -- into two substates $m=\pm 1/2$ and $m=\pm 3/2$). In both cases the population of these substates is determined by Boltzmann law. Different population of these substates provides an equilibrium polarization (at DNP) or an equilibrium alignment (at DNA). At the temperature
$'\sim 0.5$~K (and the magnetic field $\sim 2.5$~T at DNP) the values of equilibrium orientation (both polarization and alignment) are of the same order of
$\sim 0.5$~\%.

To obtain higher nuclear polarization one adds into the target substance a para\-mag\-ne\-tic admixture whose ions have an unpaired electron in the one of outer shells. At temperature and magnetic field mentioned above an ensemble of electron spins will have approximately 100 \% polarization, because the electron magnetic moment is more than the nuclear one by about three orders of magnitude. Then, the high degree of ordering of electron spins can be transmitted to nuclei by microwave irradiation.

In the case of DNA the ground states of paramagnetic ions with electron spin $S>1/2$ are split in the crystal electric field in the same way as the states of nuclei having quadrupole moments. Since the quadrupole moment of the electron shell is more than that of the nucleus by about three-four orders of magnitude, electron spins will be completely aligned. Then, again the high degree of the  electron alignment can be transmitted to the nuclei by microwave irradiation.

Thus, one may study T invariance violation with the use of five-fold correlation if DNA is realized. The suitable targets are nuclei with large quadrupole moments having low lying p-wave resonances. The most appropriate isotopes are listed in the Table~1. Note that aligned targets are of interest not only for T invariance violation in neutron-nucleus interaction but for other nuclear physics investigations as well.

\begin{table}
\tablename{~1}

\begin{center}

\begin{tabular}{|l|c|c|l|}
\hline
\parbox{1.2cm}{\centerline{isotope}} &
\parbox{3.5cm}{\strut natural\\ \strut abundance (\%)} & 
\parbox{1.2cm}{\centerline{spin}} & 
\parbox{3.5cm}{\strut p-wave resonances \\ \strut energies (eV)}\\
\hline
\strut $^{81}$Br & 49 & 3/2 & 0.9\\
$^{93}$Nb & 100 & 9/2 & 35, 42\\
$^{105}$Pd & 22 & 5/2 & 3.9, 27.6, 41.2\\
$^{115}$In & 96 & 9/2 & 6.8, 13.5, 30\\
$^{121}$Sb & 57 & 5/2 & 37.9, 55.2\\
$^{123}$Sb & 43 & 7/2 & 67\\
$^{127}$I & 100 & 5/2 & 7.5, 10.3, 13.9, 24.6\\
$^{139}$La & 100 & 7/2 & 0.74\\
\hline
\end{tabular}

\end{center}

\end{table}

To realize DNA one needs a reasonably large monocrystal containing suitable nuclei. Besides, the target material must fulfil the following requirements:

\begin{enumerate}

\item High value of quadrupole coupling constant for the nuclei under investigation: fre\-quen\-cy of nuclear quadrupole resonance (NQR) must be higher than 30 MHz.

\item High splitting of sublevels of paramagnetic impurity: frequency of electron pa\-ra\-mag\-ne\-tic resonance (EPR) must be higher than 30 GHz.

\end{enumerate}

Presently it is known about 90 isotopes with large quadrupole moments (with half-life more than 1 year) [11]. NQR frequencies are measured for more than 3000 compounds [12] and they lye in the range from $\sim 0.1$~MHz to $\sim 1000$~MHz. The EPR frequencies for these compound doped with paramagnetic admixture were not measured. Thus, to realize DNA one should choose appropriate compound with needed NQR frequencies, to incorporate paramagnetic impurity in it, to grow a monocrystal from this material and, at last, to determine EPR frequencies for paramagnetic ions in zero magnetic field. Then it will be possible to select suitable samples for the use in the experiment on T invariance violation.

For DNA ion crystal is preferred over molecular and covalent crystals. Indeed, paramagnetic ions in the former are involved in dynamic processes of a whole lattice, otherwise the paramagnetic ion can be located in a molecular or in a cluster. Besides, in ion crystal paramagnetic impurities lose their chemical "prehistory" related to original substance incorporated into compound with nuclei under investigation. In contrast, in molecular and covalent crystals paramagnetic impurity may violate regular crystal structure and behave as extrinsic impregnation involved only into low-frequency dynamic processes of the lattice. Thus, the interaction between electron spin system and nuclear spin system may become inoperative and will not serve to the transmission of high degree of ordering of electron spin system to the nuclear spin system.

Presently the FNL JINR -- ITEP collaboration performs a work on DNA realization [13]. At the first stage the LuNbO$_4$ compound was chosen as a target material. The spin of the isotope $^{175}$Lu is equal to $I=7/2$. At $T=77$~K the NQR frequencies for $^{175}$Lu in this compound are: 80.1~MHz ($\pm 1/2 \leftrightarrow\pm3/2$), 69.6~MHz ($\pm 3/2\leftrightarrow\pm 5/2$) and 108.7~MHz ($\pm 5/2\leftrightarrow\pm 7/2$) [14]. The compound LuNbO$_4$ was doped with the paramagnetic impurity Cr$^{3+}$, and the monocrystal was grown from this material. The EPR frequency in zero magnetic field was found equal to 9.7~GHz [15]. Unfortunately, this value is rather low and can provide the degree of nuclear alignment not more than (10-15)~\% at $T\sim 1$~K. Search for and synthesis of other suitable target materials are now in progress.
 
\section{Conclusion}

Investigations of time reversal invariance violation in neutron compound resonances have the advantage over  the same investigations in the other nuclear experiments. It lies in large enhancement of time reversal invariance violation effects which is analogous to the well studied enhancement of parity violation.

At simultaneous violation of space parity and time reversal invariance (PVTV) an enhancement in a p-wave resonance (structural and dynamical) may reach the value of 10$^6$. But even for such high magnitude of the enhancement the experiment on the observation of PVTV asymmetry in the interaction of polarized neutrons with polarized nuclei should be performed with the accuracy better than 10$^{-5}$ to give a more strict limit to PVTV interaction than the one following from the current data on EDM of neutron and some atoms.

In a p-wave resonance an enhancement of T invariance violation with conservation of space parity, which is due to hypothetic PCTV interactions, may be only dynamical and hardly exceeds 10$^3$. Nevertheless, strange as it may seem, an experiment on the observation of PCTV asymmetry in the interaction of polarized neutrons with aligned nuclei with the accuracy better than 10$^{-1}$ may allow to improve the current limit to PCTV interactions following from the other nuclear experiments. In the other words, even though the accuracy of the five-fold correlation experiment will be lower by 4 orders of magnitude than the one of the three-fold correlation experiment, the results of the former will be of no less interest.

Thus, surprisingly, competitive studies of PCTV effects (five-fold correlation) may be performed in much more advantageous conditions than competitive studies of PVTV effects (three-fold correlation). The reason of this situation is the following. The data on EDM give rather strong limit for PVTV forces as compared to weak PV forces being small themselves. On the other hand, the limit on PCTV forces from the EDM data is model dependent and in certain situation can be ignored. Then, intensity of PCTV forces is limited only by direct experiments and therewith is reckoned not from the level of weak PV interactions but from the level of usual strong interaction. Note, that strong interaction exceeds weak interactions at least by the factor of 10$^7$.

Thus, although the relative limits $\lambda_{PCTV} < 10^{-4}$ and
$\lambda_{PVTV} < 10^{-4}$ are roughly the same (see above Section~3), the absolute (relative to strong interaction) limit on PCTV forces is really by 7 orders of magnitude weaker than the absolute (also with respect to strong interaction) limit on PVTV forces. So even in the absence of structural enhancement an improvement of the limit on PCTV interactions may be performed at less (in some sense) expenses than the improvement of the limit on PVTV interactions.

In principle, such situation originated more than 10 years ago, just after publication of [3], where dynamical enhancement of PCTV effects in compound neutron resonances was demonstrated. Since at that moment the quantity $\lambda_{PCTV}$ was restricted by the value of 10$^{-3}$, it was inferred in [3] that the study of PCTV effects in compound resonances even at the level 10$^{-1}$ is of significance. The absence of the effect at this level was demonstrated in [16] from the data on the asymmetry of gamma-ray emission after capture of polarized p-wave neutrons.

Further progress on study of PCTV effects in neutron resonances, of which the asymmetry related to five-fold correlation $({\bf s}[{\bf k}\times{\bf I}])({\bf k}{\bf I})$ seems the most perspective, in our opinion may be provided by realization of the DNA method. It will make possible the investigation of T invariance violation with the use of the nuclei listed in the Table~1.
\bigskip

We are grateful to L.B.Pikelner for valuable discussions. The work is supported in part by RFBR (grant N 00-15-96590) and by FNPh (grant N25/IONPh-D).
\bigskip
\bigskip

\centerline{References}
\bigskip

\begin{enumerate}
 
\item P. A. Krupchitskii, Fiz. El. Chast. At. Yad., 25 (1984) 1444 (in Russian).
A.  Ya. Alexandrovich et al., Nucl. Phys. A, 567 (1994) 541.
G. E. Mitchel et al., Phys. Rep., 354 (2001) 157.
\item V. E. Bunakov, V. P. Gudkov, Nucl. Phys. A 401 (1983) 93.
\item V. E. Bunakov, Phys. Rev. Let., 60 (1988) 2250.
\item V. A. Atsarkin, A. L. Barabanov, A. G. Beda, V. V. Novitsky, NIMA, 
440 (2000) 626.
\item A. L. Barabanov, Sov. J. Nucl. Phys., 44 (1986) 755.
\item P. Herczeg, in: Parity and Time Reversal Violation in Compound Nuclear 
States and Related Topics, Eds. N. Auerbach and J. D. Bowman, World 
Scientific, Singapure, 1996, p.214.
\item M. J. Ramsey-Musolf, hep-ph/0010023 (2000).
A. Kurylov, G. C. McLaughlin, M. J. Ramsey-Musolf, Phys. Rev. D, 63 (2001) 076007. 
\item P. R. Huffman et al., Phys. Rev. C, 55 (1997) 2684.
\item V. E. Bunakov, L. B. Pikelner, Prog. Part. Nucl. Phys., 39 (1997) 337.
\item C. D. Jeffries, Dynamic Nuclear Orientation, Interscience Publishers, N.Y. - London - Sydney, 1963.
\item Bruker, Almanac, Germany, 1996, p.96.
\item G. K. Semin, T. A. Babushkina, G. G. Yakobson, Primenenie YaKR v khimii, Leningrad, Khimiya, 1972 (in Russian).
\item V. P. Alfimenkov et al., Proceedings of IX International Seminar on 
Interaction of Neutrons with Nuclei, Dubna (2001)  462.
\item A. F. Volkov et al., Fizicheskaya khimiya, 56 (1982) 1002 (in Russian).
\item A. G. Beda et al., Kristallografiya, 47 (2002) 357 (in Russian).
\item A. L. Barabanov, E. I. Sharapov, V. R. Skoy, C. M. Frankle, 
Phys. Rev. Lett., 70 (1993) 1216.

\end{enumerate}

\end{document}